\documentclass[%
reprint,
superscriptaddress,
showpacs,
amsmath,amssymb,
pra,
]{revtex4-2}
\usepackage{graphicx}
\usepackage{dcolumn}
\usepackage{bm}
\usepackage{CJKutf8}
\usepackage{color}
\usepackage{amssymb}
\usepackage{amsfonts}
\usepackage{amsthm}
\usepackage{esint}
\usepackage[colorlinks,urlcolor=blue,linkcolor=blue,citecolor=blue,anchorcolor=blue]{hyperref}
\makeatletter

\newcommand{\Rmnum}[1]{\expandafter\@slowromancap\romannumeral #1@}

\makeatother
\begin{document}

\begin{CJK*}{UTF8}{gbsn}

\preprint{APS/123-QED}

\title{Self-Accelerating Topological Edge States}


\author{Zhuo Zhang}
\affiliation{Key Laboratory for Physical Electronics and Devices, Ministry of Education, School of Electronic Science and Engineering, Xi'an Jiaotong University, Xi'an 710049, China}

\author{Yaroslav V. Kartashov}
\affiliation{Institute of Spectroscopy, Russian Academy of Sciences, Troitsk, Moscow, 108840, Russia}

\author{Milivoj R. Beli\'c}
\affiliation{College of Science and Engineering, Hamad Bin Khalifa University, 23874 Doha, Qatar}

\author{Yongdong Li}
\affiliation{Key Laboratory for Physical Electronics and Devices, Ministry of Education, School of Electronic Science and Engineering, Xi'an Jiaotong University, Xi'an 710049, China}

\author{Yiqi Zhang}
\email[Contact author: ]{zhangyiqi@xjtu.edu.cn}
\affiliation{Key Laboratory for Physical Electronics and Devices, Ministry of Education, School of Electronic Science and Engineering, Xi'an Jiaotong University, Xi'an 710049, China}

\date{\today}

\begin{abstract}
\noindent
Edge states emerging at the boundaries of materials with nontrivial topology are attractive for many practical applications due to their remarkable robustness to disorder and local boundary deformations, which cannot result in scattering of the energy of the edge states impinging on such defects into the bulk of material, as long as forbidden topological gap remains open in its spectrum. The velocity of the such states traveling along the edge of the topological insulator is typically determined by their Bloch momentum. In contrast, here, using valley Hall edge states forming at the domain wall between two honeycomb lattices with broken inversion symmetry, we show that by imposing Airy envelope on them one can construct edge states which, on the one hand, exhibit \textit{self-acceleration} along the boundary of the insulator despite their fixed Bloch momentum and, on the other hand, \textit{do not diffract} along the boundary despite the presence of localized features in their shapes. We construct both linear and nonlinear self-accelerating edge states, and show that nonlinearity considerably affects their envelopes. Such self-accelerating edge states exhibit self-healing properties typical for nondiffracting beams. Self-accelerating valley Hall edge states can circumvent sharp corners, provided the oscillating tail of the self-accelerating topological state is properly apodized by using an exponential function. Our findings open new prospects for control of propagation dynamics of edge excitations in topological insulators and allow to study rich phenomena that may occur upon interactions of nonlinear envelope topological states.
\end{abstract}

\maketitle

\end{CJK*}

\section{Introduction}

Development of new approaches for control of propagation paths, diffraction, and shape transformations of light beams is one of the most important goals of modern photonics. Localized features in light beams propagating in free space can persist over large distances only if such beams are nondiffracting and carry infinite power. In addition to rich demonstrated classes of nondiffracting beams propagating along straight trajectories~\cite{mazilu.lpr.4.529.2010} that can be associated with different coordinate systems, where it is convenient to construct them, particular attention is paid to a broad class of self-accelerating beams~\cite{berry.ajp.47.264.1979} that possess all typical features of nondiffracting states, such as the ability for self-healing, and at the same time propagate along the curved trajectory~\cite{siviloglou.ol.32.979.2007, siviloglou.prl.99.213901.2007, broky.oe.16.12880.2008, bandres.ol.33.1678.2008, ellenbogen.np.3.395.2009, liu.ol.36.1164.2011, bekenstein.prx.4.011038.2014, driben.ol.19.5523.2014, driben.ol.38.2499.2013, zhang.ol.38.4585.2013, zhang.oe.22.7160.2014}. Even though rigorous self-accelerating and non-diffracting beams in the bulk of periodic medium represented by ``static'' photonic lattices or arrays of straight waveguides~\cite{makris.ol.39.2129.2014} may not exist, it was possible to generate in such materials a class of similar self-accelerating Wannier-Stark beams~\cite{ganainy.pra.84.023842.2011, chremmos.pra.85.063830.2012}. Various other approximations to accelerating beams have been reported in static photonic lattices as well~\cite{ganainy.pra.84.023842.2011, efremidis.ol.37.2012, kominis.oe.18165.2012, chremmos.pra.85.063830.2012, lucic.pra.88.063815.2013, makris.ol.39.2129.2014, xiao.oe.22.22763.2014, qi.ol.39.1065.2014}, see also recent review~\cite{efremidis.optica.6.686.2019}. It should be stressed that all previous results on self-accelerating beams in periodic medium were reported exclusively in topologically trivial structures.

However, when a periodic medium is characterized by the nontrivial topology of its bands, it may support a different class of diffraction-free solutions that can be localized at the boundary of such medium. Such edge states are protected by the nontrivial band topology and are localized in the direction perpendicular to the edge, while remaining extended (periodic) along the edge of the material. Due to their topological protection, the edge states are immune to disorder and imperfections in the lattice, as long as disorder is not strong enough to close the topological gap. The concept of topological insulators originates from solid state physics~\cite{hasan.rmp.82.3045.2010, qi.rmp.83.1057.2011}, and it has widely extended also in photonics~\cite{lu.np.8.821.2014, ozawa.rmp.91.015006.2019, smirnova.apr.7.021306.2020, segev.nano.10.425.2021, parto.nano.10.403.2021, yan.aom.2001739.2021, zhang.nature.618.687.2023, lin.nrp.5.483.2023, yan.npjnp.1.40.2024}. Topological robustness of the edge states make them highly promising for the development of topological lasers~\cite{harari.science.359.eaar4003.2018, bandres.science.359.eaar4005.2018, bahari.science.358.636.2017, zeng.nature.578.246.2020, zhong.lpr.14.2000001.2020}, construction of topological solitons localized due to self-action~\cite{lumer.prl.111.243905.2013, mukherjee.science.368.856.2020, ablowitz.pra.96.043868.2017, ivanov.acs.7.735.2020, tang.oe.29.39755.2021, ren.nano.10.3559.2021, zhang.prl.123.254103.2019}, and design of advanced photonic structures for protected transmission of power and information~\cite{lu.chip.1.100025.2022, xu.ap.036005.2023, jalali.pra.108.040101.2023, xia.ap.6.026005.2024}. The group velocity of the edge states, even when they are unidirectional~\cite{wang.nature.461.772.2009, rechtsman.nature.496.196.2013} usually does not change upon propagation and is typically determined by the Bloch momentum of the state along the edge. It is thus generally believed that such states cannot accelerate or decelerate in the course of propagation along the edge. Moreover, in the absence of nonlinearities, topological edge states with localized envelopes exhibit diffraction along the edge.

Therefore, the natural fundamental question arises: Is it possible to construct topological edge states that would exhibit acceleration along the edge (without introducing any gradients into the underlying lattice structure~\cite{li.pra.99.053814.2019}) and can such states preserve localized features nested in them that would not undergo dispersion in the course of propagation? The answer to this question is provided in the present work, where we join the phenomenology of self-accelerating beams and topological edge states and show that the boundaries of topologically nontrivial material can support both linear and nonlinear self-accelerating topological states. Our findings are reported for valley Hall edge states forming at the domain wall in the inversion-symmetry-broken honeycomb structure. The self-accelerating property of the constructed states allows for the adjustment of their velocity and for the reversal of their direction of motion. Self-accelerating topological edge states can also recover their missing parts --- a self-healing property inherited from nondiffracting beams. We show that nonlinearity substantially affects the envelope of such self-accelerating beams. Apodized finite-power self-accelerating edge states can circumvent sharp corners of the domain wall without backscattering. Self-accelerating beams reported here are principally new two-dimensional envelope states of topological origin constructed on edge states belonging to topological gap that dictates their unusual internal phase and intensity distributions. They sharply contrast with one-dimensional Airy beams reported in trivial uniform or plasmonic media.

\section{Theoretical model}

\begin{figure}[t]
	\centering
	\includegraphics[width=\columnwidth]{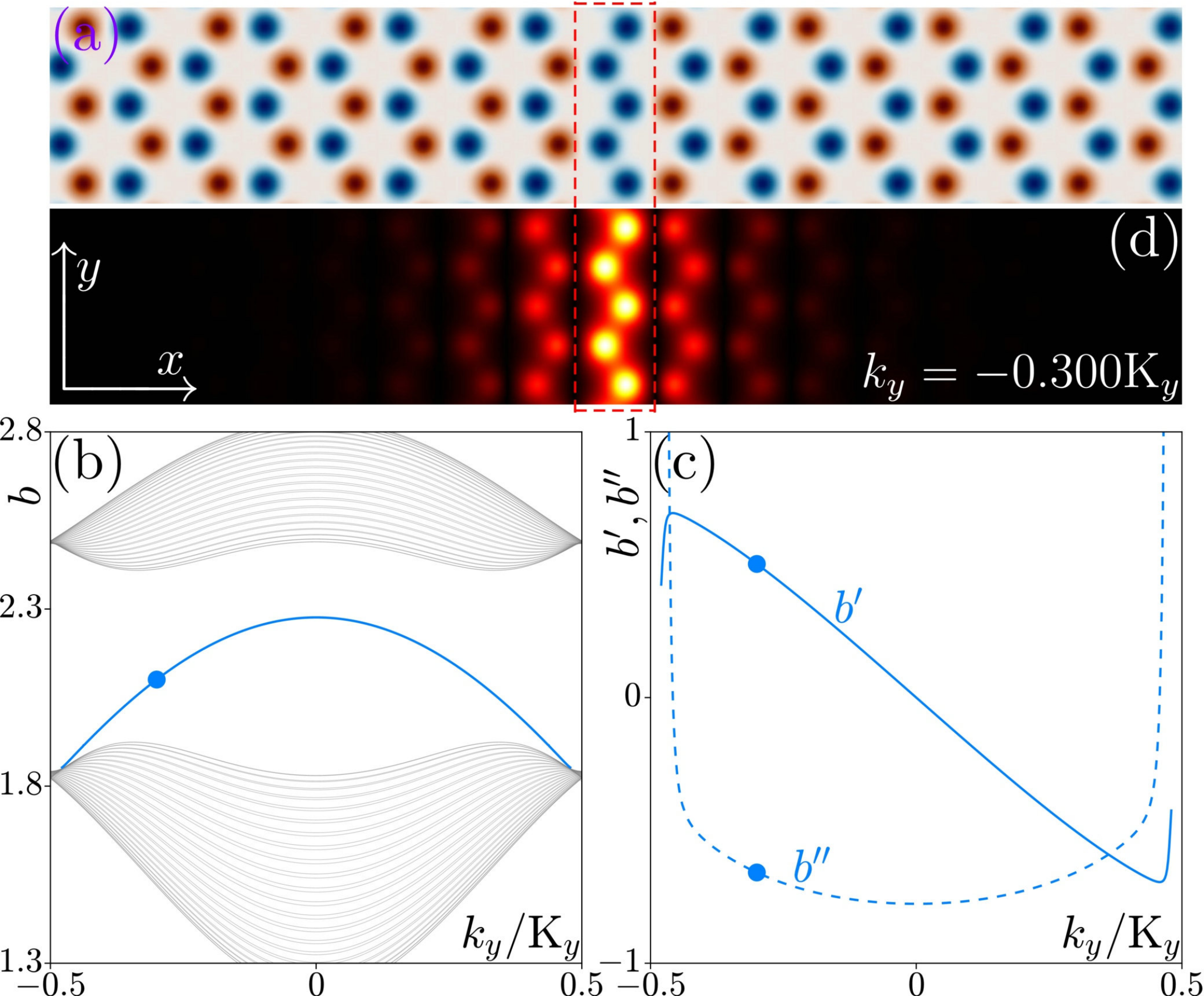}
	\caption{(a) Inversion-symmetry-broken honeycomb waveguide array with the domain wall indicated by the dashed rectangle. The depth of the red and blue waveguides is ${p+\delta}$ and ${p-\delta}$, respectively. (b) Band structure of the array from panel (a). The blue and gray lines represent propagation constants of the valley Hall edge state and of the bulk states, respectively. (c) First-order ($b'$, solid line) and second-order ($b''$, dashed line) derivatives of the propagation constant $b$ of the valley Hall edge state. (d) Field modulus distribution $|\psi|$ of the valley Hall edge state at $k_y=-0.3\textrm{K}_y$ corresponding to the blue dot in panel (b). Panels (a) and (d) correspond to ${-20\le x \le 20}$ and ${-3.5\le y \le 3.5}$ windows.}
	\label{fig1}
\end{figure}

The propagation dynamics of the light beam in the material with shallow transverse refractive index modulation defining topological waveguide array and cubic focusing nonlinearity can be described by the dimensionless Schr\"odinger equation for dimensionless field amplitude $\psi$:
\begin{equation}\label{eq1}
	i\frac{\partial \psi}{\partial z} = - \frac{1}{2} \Delta \psi - \mathcal{R}(x,y) \psi - |\psi|^2 \psi,
\end{equation}
where ${\Delta = \partial_x^2+\partial_y^2}$ is the transverse Laplacian and $(x,y)$ and $z$ are the normalized transverse coordinates and propagation distance, respectively. The function
\[
\mathcal{R}(x,y)=\sum_{m,n} p_{m,n}e^{-{[(x-x_{m,n})^2+(y-y_{m,n})^2]}/{d^2}}
\]
describes the refractive index distribution in the honeycomb waveguide array with waveguides having depths $p_{m,n}$, identical widths $d$, and located in positions with coordinates $(x_{m,n}, y_{m,n})$, see schematic illustration in Fig. \ref{fig1}. We assume that honeycomb array (frequently named ``photonic graphene'') consists of two sublattices that are detuned, i.e. ${p_{m,n}=p\pm\delta}$, with typical value of detuning $\delta=0.5$. We set here the following parameter values: Lattice constant ${a=1.6}$, waveguide width ${d=0.5}$, and depth ${p=8}$. In the arrays fabricated using direct fs-laser writing in fused silica~\cite{rechtsman.nature.496.196.2013, kirsch.np.17.995.2021, kartashov.prl.128.093901.2022, arkhipova.sb.68.2017.2023, ren.light.12.194.2023, zhong.light.13.264.2024, yan.npjnp.1.40.2024, kartashov.pu.67.1095.2024, kompanets.am.37.2500556.2025, zhang.ap.7.034002.2025} the transverse coordinates $(x,y)$ can be normalized to the characteristic transverse scale ${r_0=10\,\mu {\rm m}}$, the propagation distance $z$ is normalized to $k r_0^2\approx 1.1\,\rm mm$, with the wavenumber ${k=2n_0\pi/\lambda}$, where the background refractive index is ${n_0=1.45}$, and the wavelength in vacuum ${\lambda=800\,\rm nm}$. The waveguide depth $p$ is related to the refractive index change $\Delta n$ through ${p=k^2 r_0^2 \Delta n / n_0}$, and ${p=8}$ will result in ${\Delta n\sim8.8\times10^{-4}}$.

As mentioned above, if one sets the depth of one sublattice to ${p-\delta}$, while that of the other sublattice to ${p+\delta}$, the inversion symmetry of the array will be broken. Two such arrays with the opposite signs of detuning can be joined to created a domain wall. In Fig.~\ref{fig1}(a) we present a typical example of such a domain wall highlighted by the dashed rectangle. Note that the domain wall is periodic in $y$ with the period ${{\rm Y}=3^{1/2}a}$. It has already been demonstrated that such domain walls can support valley Hall topological edge states~\cite{wu.nc.8.1304.2017, noh.prl.120.063902.2018, tang.oe.29.39755.2021, ren.nano.10.3559.2021}, since the difference between the two valley Chern numbers of the same valley across the domain wall is $1$~\cite{zhong.ap.3.056001.2021, tang.chaos.161.112364.2022, tang.rrp.74.504.2022}.

When the array is limited along the $x$ axis, the general solution of Eq.~(\ref{eq1}) can be written as 
\[
\psi=u(x,y)\exp(i k_y y + i b z),
\]
where $b$ is the propagation constant of the edge state and $k_y$ is its Bloch momentum. Substituting this solution into Eq.~(\ref{eq1}), one obtains the equation:
\begin{equation}\label{eq2}
	{bu=\frac{1}{2} \left(\frac{\partial^2}{\partial x^2} + \frac{\partial^2}{\partial y^2} + 2ik_y \frac{\partial}{\partial y} - k_y^2\right) u + \mathcal{R} u + |u|^2 u},
\end{equation}
which in the absence of nonlinearity (i.e. when the term $|u|^2u$ is omitted) can be numerically solved to obtain the relation between $b$ and $k_y$ in the first Brillouin zone $[-{\rm K}_y/2,\, {\rm K}_y/2]$ with ${{\rm K}_y=2\pi/{\rm Y}}$ (see the details of the numerical methods in the \textbf{Appendix~\ref{appA}}). In this manner, one obtains the linear band structure of the array displayed in Fig.~\ref{fig1}(b). One finds that for this type of the domain wall, the valley Hall edge state that is shown with blue line emerges from the lower bulk band (shown with gray lines) and disappears in the same band (because the system possesses time-reversal symmetry). The first-order derivative ${b'=db/dk_y}$ and the second-order derivative ${b''=d^2b/dk_y^2}$ of propagation constant determine the group velocity and dispersion of the edge state, respectively, and are shown in Fig.~\ref{fig1}(c) by the solid and dashed curves. The valley Hall edge state moves in the negative $y$  direction when ${k_y<0}$ and in the positive $y$ direction when ${k_y>0}$. Typical valley Hall edge state at ${k_y=-0.3{\rm K}_y}$, shown in Fig.~\ref{fig1}(d) features oscillating and decaying tails at both sides of the interface reflecting its nontrivial topological nature. By changing the sign of the detuning $\delta$ one can realize the situation, when the domain wall contains only deep (red) waveguides. In this case, the valley Hall edge state will emerge from the upper bulk band and disappear in the same band, while corresponding derivatives $b'$ and $b''$ will be reversed (see \cite{ren.nano.10.3559.2021, tang.chaos.161.112364.2022}) in comparison with the values shown in Fig.~\ref{fig1}(c). We would like to note that the domain wall considered here can support bright valley Hall edge solitons, since for the sign of detuning considered here the derivative ${b''<0}$ for edge state on which such solitons can be constructed \cite{tang.oe.29.39755.2021}. The sign of detuning $\delta$ does not affect the conclusions reported below even though it may change the sign of $b''$, in which case one can construct dark valley Hall edge solitons~\cite{ren.nano.10.3559.2021}. Further we consider the $\delta=0.5$ case.

As one can see from Fig.~\ref{fig1}, both group velocity $b'$ and dispersion $b''$ are the functions of Bloch momentum $k_y$ and, therefore, if the momentum of the unconstrained (along the $y$-axis) linear edge state does not change in the course of evolution, its group velocity does not change as well. Nevertheless, our aim is to show that by imposing the proper envelope on the edge state with selected $k_y$ (importantly, such envelope should not be narrow) it is possible to realize the situation, when the features of this envelope would exhibit acceleration upon propagation.

\section{Self-accelerating envelope for the valley Hall edge states}

The valley Hall edge state displayed in Fig.~\ref{fig1}(d) is periodic in $y$, but one can superimpose a broad (in comparison with array period $\textrm{Y}$) envelope on it to construct on its basis various topological objects. Among them are localized topological edge solitons bifurcating under the action of nonlinearity from the extended edge states, whose theory for continuous media was developed in~\cite{ivanov.acs.7.735.2020, ivanov.pra.103.053507.2021, ivanov.lpr.16.202100398.2022, ren.nano.10.3559.2021, tian.fop.17.53503.2022}. Here we use a similar approach, but now we do not impose the requirement of localization on corresponding envelope. Following this method, we introduce the ansatz 
\begin{equation}\label{eq3}
	\psi = \mathcal{A}(\eta,z) u (x,y) \exp(i k_y y + i b z), 
\end{equation}
plug it into Eq.~(\ref{eq1}), and use the multiscale approach~\cite{ivanov.acs.7.735.2020} to obtain the following nonlinear Schr\"odinger equation for the envelope $\mathcal{A}$:
\begin{equation}\label{eq4}
	i \frac{\partial\mathcal{A}}{\partial z} = \frac{1}{2} {\rm sgn}(b'') \frac{\partial^2 \mathcal{A}}{\partial \eta^2} - \chi |\mathcal{A}|^2 \mathcal{A},
\end{equation}
where 
\begin{equation*}
	\eta = \frac{y+b'z}{|b''|^{1/2}}, \quad \chi=\int_{-\infty}^{+\infty}dx \int_0^{\rm Y} dy |u|^4.
\end{equation*}
is the transverse coordinate running with group velocity $-b'$ of the edge state, and effective nonlinear coefficient determined by the shape of the edge state, respectively. Here we use the standard normalization condition $\int_{-\infty}^{+\infty}dx \int_0^{\rm Y} dy |u|^2=1$ for linear edge state.

The Eq.~(\ref{eq4}) possesses self-accelerating self-trapped solutions~\cite{kaminer.prl.106.213903.2011} exhibiting parabolic trajectories that represent nonlinear generalizations of the Airy beams. To obtain them, one can move into accelerating coordinate frame $\eta \to{\eta-\mu z^2}$ that yields the equation:
\begin{equation}\label{eq5}
	{i \frac{\partial \mathcal{A}}{\partial z} = i 2\mu z \frac{\partial \mathcal{A}}{\partial \eta} - \frac{1}{2} \frac{\partial^2 \mathcal{A}}{\partial \eta^2} - \chi |\mathcal{A}|^2 \mathcal{A}},
\end{equation}
where  $\mu$ is a free parameter determining the parabolic trajectory. Assuming that the solution of this equation can be written in the form: 
\begin{equation}\label{eq6}
	\mathcal{A}(\eta,z) = \frac{w(\eta)}{\sqrt{\chi}} \exp \left[ i\left(b_{\rm nl}z + 2\mu \eta z + \frac{2}{3} \mu^2 z^3 \right) \right],
\end{equation}
one arrives at the ordinary differential equation for beam profile $w(\eta)$:
\begin{equation}\label{eq7}
	\frac{\partial ^2 w}{\partial \eta^2} + 2 |w|^2w - 4\mu \eta w - 2b_{\rm nl} w=0.
\end{equation}
This equation already does not contain parameters $\chi$, $b'$, and $b''$ depending on the momentum $k_y$ of the valley Hall edge state and can therefore be used to obtain envelopes for any $k_y$ value. Notice that the propagation constant $b_\textrm{nl}$ introduced in Eq.~(\ref{eq6}) by analogy with propagation constant of self-sustained states propagating along the straight trajectories, can now be eliminated by shifting the solution $w(\eta)$ by $b_{\rm nl}/(2\mu)$ in $\eta$. Thus, it makes sense to compare phase accumulation rate arising due to nonlinearity only for beams with global intensity maximum in the same transverse location $\eta$. Such nonlinear phase accumulation rate can be determined at the initial stages of propagation of the self-accelerating edge state [where cubic contribution $\sim (2/3)\mu^2z^3$ to phase arising due to propagation along curved path is still small for $\mu \ll 1$] using the expression
\begin{equation}\label{eq8}
	b_{\rm nl} \approx \frac{\arg \langle \mathcal{A}(\eta,z),\mathcal{A}(\eta,z+\Delta z) \rangle}{\Delta z} - 2\mu \eta,
\end{equation}
where $\Delta z$ should be small. We therefore further calculate $b_\textrm{nl}$ for beams with global intensity maximum located at $\eta=0$. Due to this requirement, the phase shift calculated using Eq.~(\ref{eq8}), may approach some small nonzero value $b_0$ when $|\mathcal{A}|_{\max} \to 0$, so it is more convenient to plot the quantity ${\beta=b_{\rm nl}-b_0}$ as a propagation constant or ``energy shift'' of self-accelerating nonlinear state. The so defined quantity is independent of the $\eta$-location of global maximum of the beam.

\section{Self-accelerating valley Hall edge states}
\subsection{Linear case}

In the following, we construct both linear and nonlinear self-accelerating valley Hall edge states by superimposing the envelope obtained from Eqs.~(\ref{eq6}) and (\ref{eq7}) onto exact valley Hall edge states and studying their propagation dynamics. If the nonlinear term in Eq.~(\ref{eq7}) is omitted, one can obtain the following explicit solution in the form of linear Airy beam:
\begin{equation}\label{eq9}
	w(\eta)={\rm Ai} [(4\mu)^{1/3} \eta].
\end{equation}
Here, we used the variable $\eta$ instead of ${\eta+b_{\rm nl}/(2\mu)}$, since nonlinearity-induced phase shift is irrelevant in this case. We superimpose such Airy envelope onto the valley Hall edge state at ${z=0}$, to obtain the initial field distribution ${\psi(x,y)={\mathcal A}(y) u(x,y) \exp(ik_yy)}$ and model its propagation dynamics in real waveguide array using linear version of Eq. (\ref{eq1}). The so-constructed input represents \textit{hybrid state} that is localized \textit{across} the domain wall due to its topological nature, and at the same time having localized features \textit{along} the domain wall due to oscillations present in the shape of Airy function (the power of the beam still remains infinite because oscillating tail does not decay exponentially). We adopt here sufficiently small value of parameter ${\mu=0.002}$ to ensure that the main lobe of the so-constructed state is sufficiently wide, so that the envelope equation (\ref{eq4}) and multiscale approach are applicable. Because the frequency of oscillations on the tail of Airy beam gradually increases, the validity of this approximation may sooner or later be violated, but usually this happens very far from the global beam maximum (in the region, where the amplitude of the beam becomes very small) and arising distortions do not notably affect beam evolution.

\begin{figure}[t]
	\centering
	\includegraphics[width=\columnwidth]{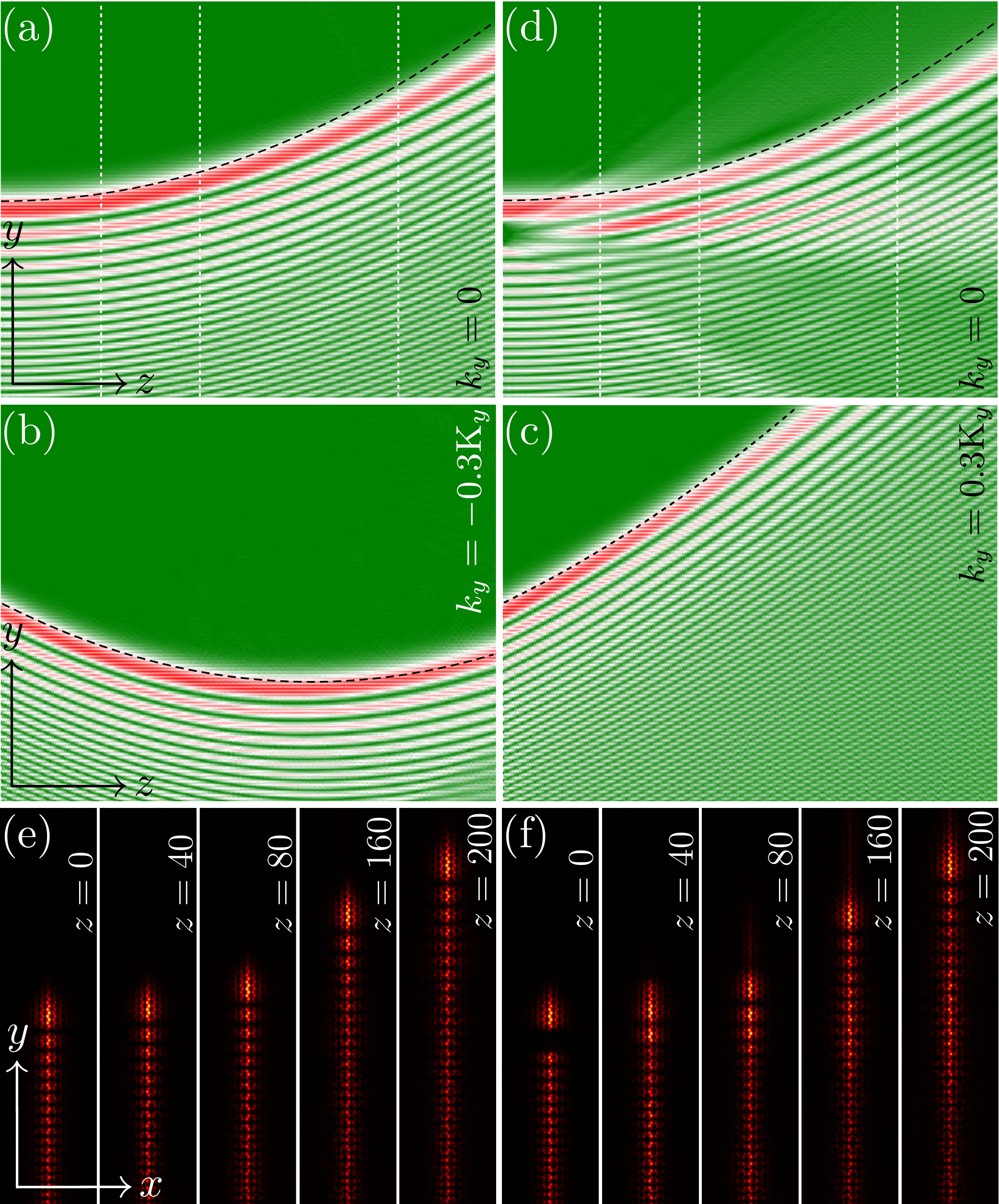}
	\caption{(a) Cross-section ${|\psi(x=0,y)|}$ illustrating propagation dynamics of the valley Hall edge state with ${k_y=0}$ and superimposed Airy envelope with ${\mu=0.002}$. The parabolic dashed line is the theoretically predicted trajectory of the self-accelerating valley Hall edge state. The dynamics is shown within the window ${0\le z \le 200}$ and ${-80\le y \le 80}$. (b,c) Same as in (a), but for the valley Hall edge states with Bloch momenta ${k_y=-0.3{\rm K}_y}$ and ${k_y=0.3{\rm K}_y}$, respectively. (d) Self-healing of the self-accelerating valley Hall edge state from (a) with eliminated second lobe. (e) Field modulus distributions $|\psi(x,y)|$ at distances corresponding to the vertical dashed lines in (a) that clearly illustrate self-acceleration of the beam along the domain wall. Panels (e) are shown within the window ${-20 \le x \le 20}$ and ${-80\le y \le 80}$. (f) Field modulus distributions at different distances $z$ corresponding to the vertical dashed lines in (d).}
	\label{fig2}
\end{figure}

In Fig.~\ref{fig2}(a) we illustrate propagation dynamics of the self-accelerating beam constructed on the valley Hall edge state with $k_y=0$. The group velocity of carrier edge state ${v=-b'}$ is zero for this momentum value, so such edge state with usual localized Gaussian envelope would not move and would exhibit diffraction (see the \textbf{Appendix~\ref{appB}}). Nevertheless, the presence of the asymmetric Airy envelope immediately leads to self-acceleration of the beam along the domain wall of topological insulator with $z$ (akin to self-acceleration exhibited by usual Airy beams in free space \cite{siviloglou.ol.32.979.2007}). This illustrates that even though the momentum of the carrier edge state is well defined at all propagation distances, thereby determining the velocity of the \textit{carrier state}, its envelope can still shift along the domain wall with different and varying with $z$ velocity and namely the latter velocity determines the shift of localized features in beam profile. 
The reason is that when superimposing an Airy envelope (with its characteristic asymmetric momentum spectrum) onto the original valley edge state, the momentum distribution of the resulting state reflects a convolution of both components. Thus, even though the original edge state has zero group velocity, the Airy envelope would shift the momentum components of the modulated valley Hall edge state predominantly to ${k_y>0}$.
The dashed curve superimposed on Fig.~\ref{fig2}(a) corresponds to the expected from the envelope equation ${y=|b''|^{1/2}\mu z^2}$ trajectory of self-accelerating beam and it is indeed close to the actual trajectory obtained by simulating beam propagation in the original Eq. (\ref{eq1}) (the deviation is expected to come from reshaping of the envelope that is unavoidable due to higher-order derivatives that are neglected in the envelope equation). The self-accelerating valley Hall edge state maintains its profile over sufficiently large propagation distances, with the width of the main and subsequent lobes and structure of the beam remaining nearly unchanged (i.e., illustrating non-diffracting propagation), as it is evident from distributions in the $(x,y)$ plane shown at different distances in Fig.~\ref{fig2}(e), which correspond to the vertical dashed lines in Fig.~\ref{fig2}(a). The progressively increasing shift of the beam (acceleration) is also obvious from these plots. One can also see that radiation into the bulk is absent due to topological nature of the state.

Since the imposed Airy envelope forces valley Hall edge state to accelerate in the positive $y$ direction, one may assume that if the carrier state initially moves in the negative direction of the $y$ axis, its propagation direction can be reversed with $z$ due to the impact of the envelope, while for the state initially moving in the positive direction of the $y$ axis, the acceleration will further increase the initial velocity. In Fig.~\ref{fig2}(b) we show propagation dynamics of the self-accelerating valley Hall edge state with momentum ${k_y=-0.3{\rm K}_y}$, corresponding to negative group velocity $-b'$ of the carrier state. This state initially indeed moves in the negative direction of the $y$-axis, but then changes its propagation direction when acceleration due to the imposed envelope changes the sign of velocity. Remarkably, this state still evolves practically without changing its envelope. This phenomenon, demonstrated previously for Airy beams in free space~\cite{siviloglou.ol.33.207.2008, zhang.rrp.67.1099.2015}, has never been reported in topological insulators, where it is commonly believed that in the absence of defects or gradients the edge states cannot change their propagation direction. If the carrier edge state with Bloch momentum ${k_y=+0.3{\rm K}_y}$ corresponding to positive group velocity $-b'$ is used for the construction of self-accelerating state, then one observes progressively increasing with distance $z$ displacement of features of Airy envelope demonstrated in Fig.~\ref{fig2}(c). We note that the peak amplitude of the beam in Figs.~\ref{fig2}(a) and \ref{fig2}(c) slightly reduces during propagation, while in Fig.~\ref{fig2}(b) it changes only weakly, at least at the propagation distance shown. This is the consequence of slight reshaping of the beam upon its propagation along the domain wall (because the imposed envelope does not take into account the presence of higher-order derivatives that were neglected in the envelope equation, see the explanation above). Another reason is that in simulations we use very large, but finite $y$-windows, so in contrast to ideal Airy beam that has long slowly decaying oscillating tail, our input beam is truncated by the window far away from the main lobe and this may also lead to slow power transfer from the main lobe into tails due to self-healing tendency. This transfer is just delayed in Fig. \ref{fig2}(b) leading to practical invariance of amplitude with $z$. 
Note that the reversal of the propagation direction in Fig.~\ref{fig2}(b) and enhanced acceleration in Fig.~\ref{fig2}(c) correspond to distinct modifications of the momentum distributions around their original $k_y$ values due to superposition of the envelopes.
The theoretical prediction ${y = -b'z + |b''|^{1/2}\mu z^2}$ for beam propagation trajectory shown with dashed lines in Figs.~\ref{fig2}(b) and \ref{fig2}(c), where ${b'\sim \pm 0.5029}$, respectively, is in reasonable agreement with actual trajectory obtained on the basis of simulations of Eq.~(\ref{eq1}).

One of the most distinguishing features of non-diffracting beams, including accelerating Airy beams, is their ability to self-heal from localized introduced perturbations~\cite{broky.oe.16.12880.2008}. This property is a consequence of non-diffracting nature of corresponding beams and infinite power that they carry under ideal conditions. Physically, when the localized perturbation is imposed on the beam, it rapidly diffracts in the course of evolution, while the beam remains unaffected, so that after sufficiently long distance $z$ one observes visually the recovery of the ideal beam shape. We confirmed that this property also holds for self-accelerating valley Hall edge states. The state in Fig.~\ref{fig2}(d) with removed second lobe indeed self-heals upon propagation, while the trajectory of its motion remains practically unaffected by the introduced disturbance [compare Fig. \ref{fig2}(d) with Fig. \ref{fig2}(a)]. The field modulus distributions at different distances illustrating recovery of the second lobe that was removed at $z=0$ are presented in Fig.~\ref{fig2}(f). The comparison of distributions in Figs.~\ref{fig2}(e) and \ref{fig2}(f) also demonstrates that the internal structure of the beam is recovered after sufficiently large propagation distance.

It should be stressed that the envelope theory leading to Eq. (\ref{eq4}) requires slow variation of the envelope of the beam $\mathcal{A}$ on one period $\textrm{Y}$ of the domain wall, while increasing $\mu$ reduces the scale of the characteristic features in Airy beam envelope and simultaneously leads to faster bending of the beam. Thus, the validity of the envelope theory requires small values of $\mu$ and increase of this parameter would lead to more pronounced deviations of actual propagation trajectory from parabolic one (and more pronounced reshaping, especially on the oscillating tails of the beam). A similar conclusion was obtained for approximations of the non-diffracting beams in trivial lattices~\cite{ganainy.pra.84.023842.2011, chremmos.pra.85.063830.2012}. Nevertheless, we were able to see self-acceleration of the edge states even for $10$ times larger values of $\mu \sim 0.02$ indicating on robustness of the phenomenon. In addition, we found that self-accelerating properties persist, at least at the initial stages of propagation, even if the beam artificially apodized with a Gaussian envelope (see the section on topological protection). Thus we, for the first time to our knowledge, presented self-accelerating, non-diffracting and self-healing topological states.

It is worth noting that the trajectory of the accelerating waves can be not only parabolic, see for instance an example proposed in previous literature~\cite{froehly.oe.19.16455.2011,yan.jo.16.035706.2014}. Such envelopes can be also used to produce self-accelerating valley Hall edge states with different from parabolic trajectories. We however, would like to leave the investigation of such envelopes for the future studies and focus more on the simplest Airy envelope that leads to acceleration along the parabolic trajectory.

\subsection{Nonlinear case}

We now take into account nonlinearity of the material and obtain nonlinear generalizations of self-accelerating valley Hall edge states. To calculate the envelope, we use Eq. (\ref{eq7}) with cubic nonlinear term and obtain its solutions using shooting method, assuming that at sufficiently large positive values of $\eta$, where the envelope function $w(\eta)$ decays exponentially, the nonlinear term can be omitted and the asymptotic values of the function and its first derivative are given by ${w(\eta) = \sigma {\rm Ai}(\eta)}$ and ${w'(\eta) = \sigma {\rm Ai}'(\eta)}$, where $\sigma$ is the free parameter that can be tuned to adjust the position of the main lobe of the beam (that we require to be located at $\eta=0$). It is known from theory of topological edge solitons \cite{smirnova.apr.7.021306.2020, szameit.np.20.905.2024} that the nonlinearity shifts the propagation constant of the nonlinear edge state from corresponding linear eigenvalue, so that the nonlinear state may enter into the band and couple with the bulk states, thereby losing its localization. Therefore, when we calculate the family of nonlinear Airy-like envelopes we track the ``energy shift'' $\beta=b_\textrm{nl}-b_0$ [see Eq.~(\ref{eq8})] as a function of peak amplitude of the edge state to compare it with the width of the gap to avoid coupling of such nonlinear self-accelerating edge states with bulk modes.

\begin{figure}[t]
\centering
\includegraphics[width=\columnwidth]{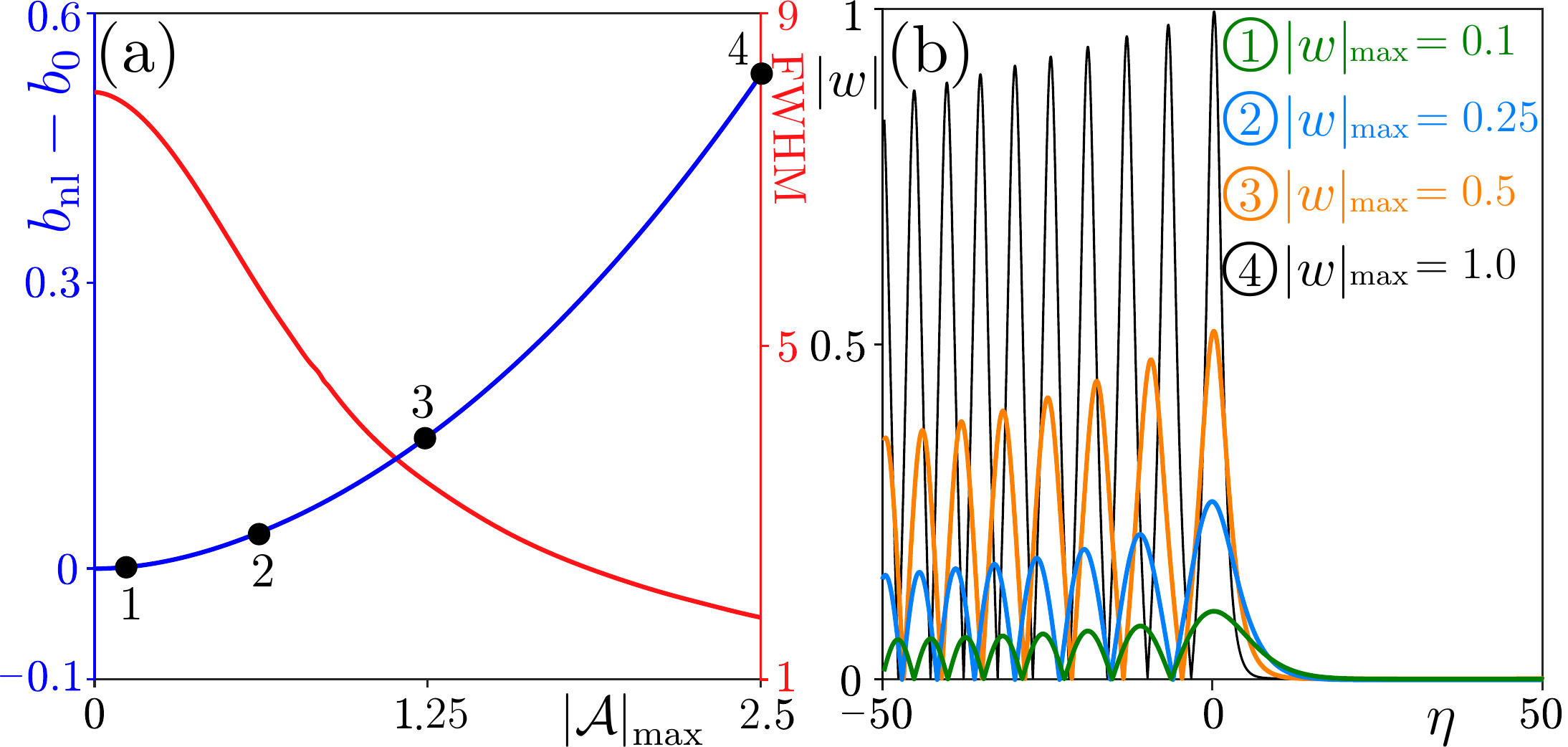}
\caption{(a) Blue curve (ref. the left $y$ axis): Peak amplitude of the nonlinear self-accelerating beam versus energy shift $\beta=b_\textrm{nl}-b_0$. Red curve (ref. the right $y$ axis): FWHM of the main lobe in the intensity distribution of the nonlinear self-accelerating solution versus its peak amplitude with ${k_y=0}$. The ``energy shift'' corresponding to the dots labeled ${1\sim4}$ is given by $0.006$, $0.038$, $0.138$, and $0.516$, respectively. (b) Profiles of the self-accelerating solutions for different peak amplitudes $|w|_{\max}$, corresponding to the dots in (a).	For all cases: ${\mu=0.002}$.}
\label{fig3}
\end{figure}

\begin{figure}[t]
	\centering
	\includegraphics[width=\columnwidth]{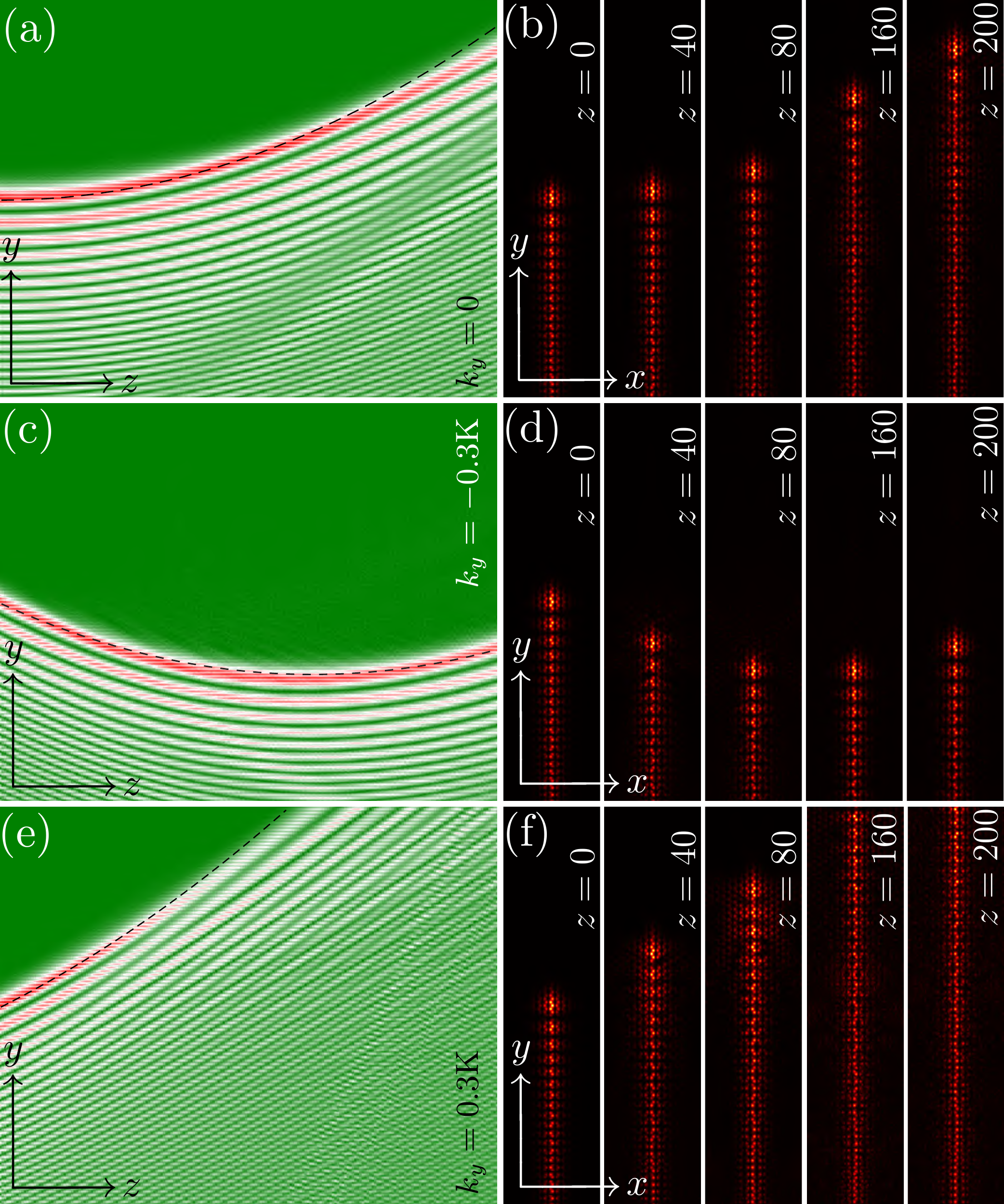}
	\caption{(a) Evolution dynamics of the nonlinear self-accelerating valley Hall edge state at ${k_y=0}$, for ${\chi\approx0.1592}$, ${b''\approx-0.7763}$, ${\mu=0.002}$, and ${|\mathcal{A}|_{\max}=0.63}$. (b) Field modulus distributions $|\psi(x,y)|$ at selected propagation distances. (c,d) Same as in (a,b), but for ${k_y=-0.3{\rm K}_y}$, at ${\chi\approx0.1663}$, ${|b'|\approx0.5029}$, ${b''\approx-0.6584}$, and ${|\mathcal{A}|_{\max}=0.61}$. (e,f) Same as (c,d) but for and ${k_y=+0.3{\rm K}_y}$. Dashed lines in (a,c,e) stand for the predicted accelerating trajectories. Panels in (a,c,e) are shown in the window ${0\le z \le 200}$, ${-80\le y \le 80}$. Panels in (b,d,f) are shown in the window ${-20 \le x \le 20}$ and ${-80\le y \le 80}$. For all cases: ${\mu=0.002}$.}
	\label{fig4}
\end{figure}

In Fig.~\ref{fig3}(a), we display the ``energy shift'' (blue curve) as well as the full width at half maximum (FWHM) of the first lobe in the intensity distribution of the beam (red curve) as functions of the peak amplitude ${|\mathcal{A}|_{\max}=|w|_\textrm{max}/\chi^{1/2}}$ for nonlinear self-accelerating solutions with ${\mu=0.002}$ and ${k_y=0}$ (the curve for ${k_y=\pm0.3{\rm K}_y}$ is quite similar). One finds that increasing peak amplitude leads to narrowing of all lobes in the profile of the beam and growing ``energy shift''. To illustrate the transformation of the envelope, we show in Fig.~\ref{fig3}(b) the envelopes corresponding to the dots in Fig.~\ref{fig3}(a). Note that with increasing peak amplitude, the widths of different lobes gradually equilibrate. Because the difference between the top edge of the gap and the propagation constant of the edge state depends on momentum $k_y$, one should compare this difference with nonlinear ``energy shift'' for different $k_y$ values to ensure that nonlinear self-accelerating edge state will be located in the gap. The interval between the eigenvalue of linear valley Hall edge state and top edge of the gap is about $0.31$ for ${k_y=\pm0.3{\rm K}_y}$ and about $0.21$ for ${k_y=0}$. Therefore, the envelopes corresponding to the dots ${1\sim3}$ in Fig.~\ref{fig3}(a) correspond to nonlinear edge states with propagation constants in the topological gap, while the state with envelope corresponding to the dot ${4}$ is in the bulk band.

\begin{figure*}[t]
\centering
\includegraphics[width=\textwidth]{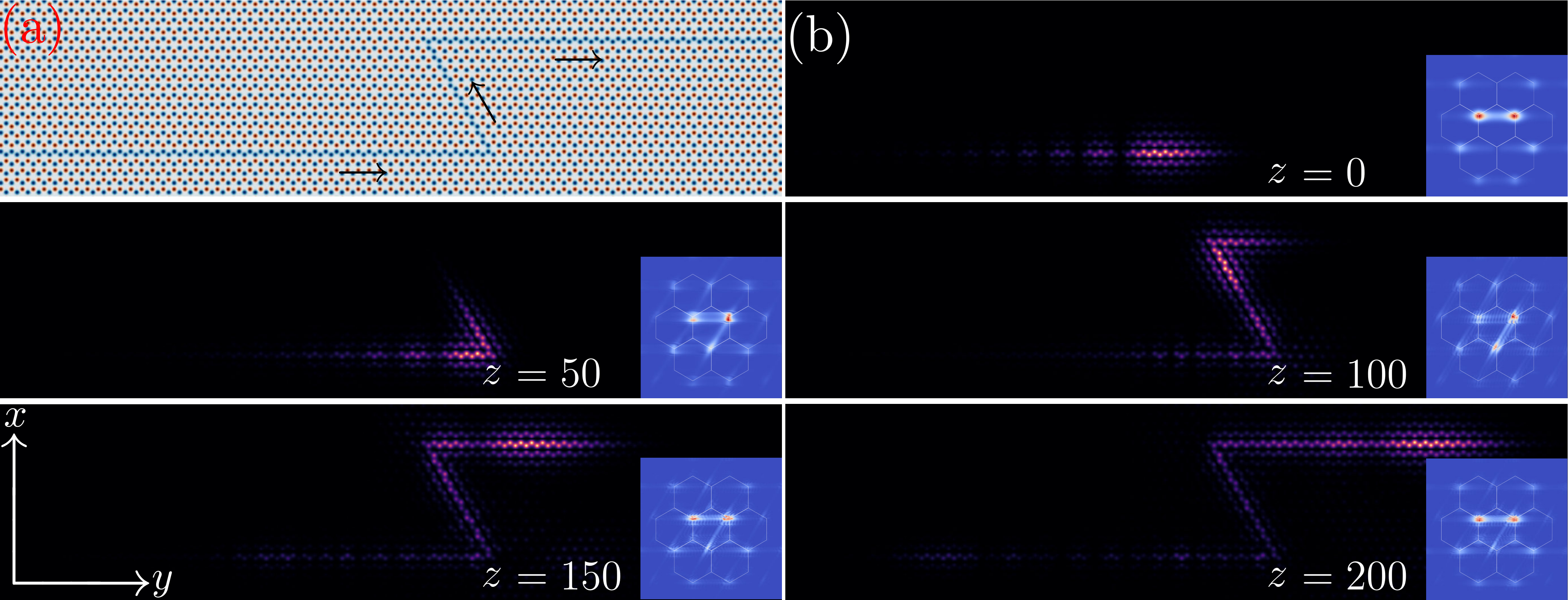}
\caption{(a) Composite photonic graphene lattice with an $Z$-path domain wall indicated by the blue color. The arrows indicate the propagation direction of the input beam. (b) Field modulus distributions of a finite-energy self-accelerating valley Hall edge state at different distances illustrating passage through the $Z$-shaped region. The insets show the spatial spectrum of the beam in the Fourier domain with hexagons representing Brillouin zone. All panels are shown within the window  ${-20 \le x \le 20}$, ${-100\le y \le 100}$. The insets are shown within the window ${-5 \le k_{x,y} \le 5}$. 
}
\label{fig5}
\end{figure*}

To test robustness of propagation in nonlinear case we prepared the nonlinear self-accelerating valley Hall edge state with envelope corresponding to peak amplitude ${|w|_{\max}=0.25}$ (or ${|\mathcal{A}|_{\max}=|w|_\textrm{max}/\chi^{1/2}}$) and superimposed calculated envelope on the linear carrier edge state. In Fig.~\ref{fig4}(a) we illustrate the propagation dynamics of such state at ${k_y=0}$ in the frames of the original Eq. (\ref{eq1}). The field modulus distributions at different selected propagation distances are shown in Fig.~\ref{fig4}(b). The results clearly demonstrate self-acceleration of the state in the course of propagation. The propagation dynamics of nonlinear self-accelerating states with ${k_y=\pm0.3{\rm K}_y}$ are shown in Figs.~\ref{fig4}(c,d) and Figs.~\ref{fig4}(e,f), respectively. The state corresponding to ${k_y=-0.3{\rm K}_y}$ shows somewhat more stable evolution practically without modifications of the envelope in comparison with ${k_y=+0.3{\rm K}_y}$ beam. One can conclude that self-accelerating edge states persist even in the presence of nonlinearity of the medium. 

Finally, we note that the nonlinear self-accelerating states exist not only in the focusing medium, but also in defocusing one, by analogy with topological solitons \cite{zhong.pra.107.L021502.2023}. The example of the envelope of the nonlinear edge state in defocusing medium and its propagation dynamics are presented in the \textbf{Appendix~\ref{appD}}. Along the same lines, nonlinear self-accelerating valley Hall edge states can also be constructed in media with saturable nonlinearity~\cite{zhang.ol.38.4585.2013, zhang.oe.22.7160.2014} typical for photorefractive crystals~\cite{zhong.ap.3.056001.2021}, i.e. they are rather universal.

\subsection{Topological protection}

The most representative manifestation of the topological protection of the edge states in valley Hall systems, including valley Hall edge solitons~\cite{tang.oe.29.39755.2021, ren.nano.10.3559.2021}, is that they can circumvent sharp corners without backward reflection or radiation into the bulk. To prove that such a protection takes place also for self-accelerating edge states, we designed here a $Z$-shaped domain wall depicted in Fig.~\ref{fig5}(a), that allows to demonstrate such a behavior. Considering that the self-accelerating valley Hall edge state at ${k_y=0.3{\rm K}_y}$ always moves in the positive direction of the $y$-axis [see Fig.~\ref{fig4}(e)], we select namely such state for illustration of such a protection.

As is well known, in valley Hall systems only the states populating valleys of the same type are topologically protected [in the first Brillouin zone, the $\bf K$ valleys are located at ${(\pm1/3^{1/2},1/3){\rm K}_y}$ and ${(0,-2/3){\rm K}_y}$, while the ${\bf K}'$ valleys are located at ${(\pm1/3^{1/2},-1/3){\rm K}_y}$ and ${(0,2/3){\rm K}_y}$]. To clearly capture the passage of the self-accelerating beam through $Z$-shaped region at the domain wall and to be sure that backward reflection is absent, we superimposed the exponential function $\exp(0.04y)$ on the self-accelerating valley Hall edge state. In Fig.~\ref{fig5}(b), we show the initial field modulus distribution of such apodized self-accelerating valley Hall edge state, while the inset in this figure shows spatial spectrum of the beam confirming that only $\bf K$ valleys were excited and that the spectrum is well-localized around corresponding valleys. When the beam reaches ${z=50}$, it circumvents the first sharp corner, while at propagation distance ${z=100}$ it circumvents the second corner. At ${z=200}$, the largest part of the beam has passed through the $Z$-shaped region. Importantly, the beam keeps propagating along the domain wall, while maintaining its Airy-like envelope (with clearly resolvable oscillations), even though corresponding lobes gradually broaden (we attribute this slow shape transformation to the apodization of the input beam). The insets with spatial spectrum demonstrate the absence of the inter-valley scattering, since the beam occupies only $\bf K$ valleys at all propagation distances. The absence of backscattering is also obvious from spatial field modulus distributions. Upon further propagation such beam will eventually evolve into Gaussian-like distribution due to its finite input power (similar transformation in trivial medium is illustrated in \cite{siviloglou.ol.32.979.2007}). At the same time, our investigation demonstrates that too long tail affects the self-accelerating edge state in the inverted space --- the longer is the tail of the edge state, the wider is the initial spectrum (it exhibits a stripe-like distribution that may extend away from the $\textbf{K}$ valleys due to rapid oscillations on the tail of the beam far away from its main lobe). Such an expansion of spectrum may eventually lead to excitation of the $\textbf{K}'$ valleys.

\section{Conclusion and Outlook}

In this work, both linear and nonlinear self-accelerating topological valley Hall edge states are predicted and analyzed. If the characteristic features of the envelope that is superimposed onto the topological edge state are sufficiently broad, the self-accelerating topological edge states can be constructed that preserve their shapes in the course of propagation, just like nondiffracting beams, but also accelerate along the domain wall. The self-accelerating topological edge states may reverse the direction of their motion during propagation. In addition to the topological protection, the self-accelerating topological edge states can also self-heal themselves if they are partially obstructed. Our study thereby connects the two previously independent fields --- the self-accelerating beams and the topological edge states. It may inspire new ideas and realizations in cold atoms, acoustics, nonlinear physics, quantum optics, and micro/nano materials. Self-accelerating beams reported here can be potentially realized in waveguide arrays fabricated by the fs direct laser writing in dielectrics or in exciton-polariton systems~\cite{kartashov.prl.119.253904.2017, banerjee.prl.124.063901.2020}.

The study performed here highlights the power of the envelope physics applied to topological edge states. Namely, constructing different types of topological objects on the edge states allows to study nontrivial transformations/evolution dynamics of their envelopes and their interactions in topological materials. This concept can be interesting not only from the point of view of nonlinear topological materials~\cite{smirnova.apr.7.021306.2020, szameit.np.20.905.2024}, but also for non-Hermitian~\cite{parto.nano.10.403.2021, yan.nano.2247.2023, nasari.ome.13.870.2023}, quantum~\cite{yan.aom.2001739.2021, hashemi.aplp.10.010903.2025}, and programmable topological photonics~\cite{capmany.nm.23.874.2024}. Our results on self-accelerating topological states can be extended to other types of the beams with different envelopes~\cite{efremidis.ol.37.2012,efremidis.pra.89.023841.2014} and potentially to non-paraxial settings~\cite{kaminer.prl.108.163901.2012, aleahmad.prl.109.203902.2012, zhang.prl.109.193901.2012, courvoisier.ol.37.1736.2012, bandres.njp.15.013054.2013}
since valley Hall edge states have been well addressed in such settings~\cite{xue.apr.2.2100013.2021,liu.aipx.6.1905546.2021}.

\renewcommand\thefigure{A\arabic{figure}} 
\setcounter{figure}{0}

\appendix
\section*{Appendices}

\section{Numerical methods}  \label{appA}

\subsection*{The plane-wave expansion method.}	
By inserting the ansatz ${\psi=u(x,y)\exp(ik_y y +ibz)}$ into Eq.~(\ref{eq1}), one obtains Eq.~(\ref{eq2}). We use the plane-wave expansion method to solve Eq.~(\ref{eq2}) by neglecting the nonlinear term (that transforms this equation into linear eigenvalue problem). To solve it we expand $u$ and $\mathcal{R}$ into the Fourier series in a supercell with the sufficient number of harmonics:
\begin{equation}\label{eqs4}
\begin{split}
	u = & \sum_{m,n}c_{m,n} \exp\left(iK_m x+iK_n y\right),\\
	\mathcal{R} = & \sum_{l,s}v_{l,s} \exp\left(iK_l x+iK_s y\right)
\end{split}
\end{equation}
where $c_{m,n}$ and $v_{l,s}$ are the Fourier coefficients, $K_{m,l}=2(m,l)\pi/D_x$, $K_{n,s}=2(n,s)\pi/D_y$, $D_{x,y}$ are the sizes of the supercell along the $x,\,y$ axes, and $(m,n,l,s)$ are the integers. Due to periodicity of the system in the $y$ direction, the $D_y$ size of the supercell can be selected equal to the $\textrm{Y}$ period. Plugging the above series into the linear version of Eq.~(\ref{eq2}), after simple algebraic transformations one obtains a series of linear equations with different $(m,n,l,s)$:
\begin{equation}\label{eqs5}
-\frac{1}{2}\left[K_m^2+(K_n+k_y)^2\right] c_{m, n}+\sum_{l,s} v_{l,s} c_{m-l, n-s}= b c_{m, n}
\end{equation}
Rewriting Eq.~(\ref{eqs5}) in matrix format and diagonalizing the matrix, one obtains the eigenvalues $b$ for a given $k_y$ (i.e. the spectrum) and the corresponding eigenvectors $c_{m,n}$ that allow to construct the eigenmodes $u$ of the array according to Eq.~(\ref{eqs4}).

\subsection*{The beam propagation method.}	
To model the propagation of the beam we rewrite the Eq.~(\ref{eq1}) into 
\begin{equation}\label{eqs6}
\frac{\partial \psi}{\partial z}= \mathcal{L} \psi + \mathcal{N} \psi
\end{equation}
with ${\mathcal{L} = (i/2) (\partial_x^2 + \partial_y^2)}$ and ${\mathcal{N} = i (\mathcal{R} + |\psi|^2)}$ being linear diffraction and nonlinear operators, respectively. For small propagation steps, one can treat/apply linear and nonlinear operators successively at each propagation step. For instance, applying the Fourier transform to $\mathcal{L}\psi$ one obtains
\[
\mathcal{F} \{\mathcal{L}\psi \} = -\frac{i}{2} \left( \omega_x^2+\omega_y^2 \right) \hat{\psi}, 
\]
where $\hat{\psi}$ is the Fourier transform of $\psi$, $\omega_{x,y}$ are the frequencies. This allows to obtain complex field amplitude in Fourier domain on the next step $dz$ as
\begin{equation}
\hat{\psi}(z+dz) = \exp\left[-\frac{i}{2} \left(\omega_x^2+\omega_y^2 \right) dz \right] \hat{\psi} (z)
\end{equation}
By taking inverse Fourier transform and applying the nonlinear operator one eventually obtains
\begin{equation}
\psi(z+dz) = \exp\left(\mathcal{N}dz\right) \mathcal{F}^{-1} \left\{ \hat{\psi}(z+dz) \right\}
\end{equation}
where $\mathcal{F}^{-1}$ is the inverse Fourier transform operator.

\section{Edge state with a Gaussian envelope}\label{appB}

To illustrate that the group velocity of the edge state with simple Gaussian envelope is determined by the $-b'$ (that implies zero group velocity at $k_y=0$) we demonstrate here the dynamics for the edge state with sufficiently broad envelope $\exp(-y^2/25)$. The width of Gaussian envelope is selected here such as to be equal to the width of the first lobe of the Airy envelope used in Fig.~\ref{fig2}.

\begin{figure}[h]
\centering
\includegraphics[width=\columnwidth]{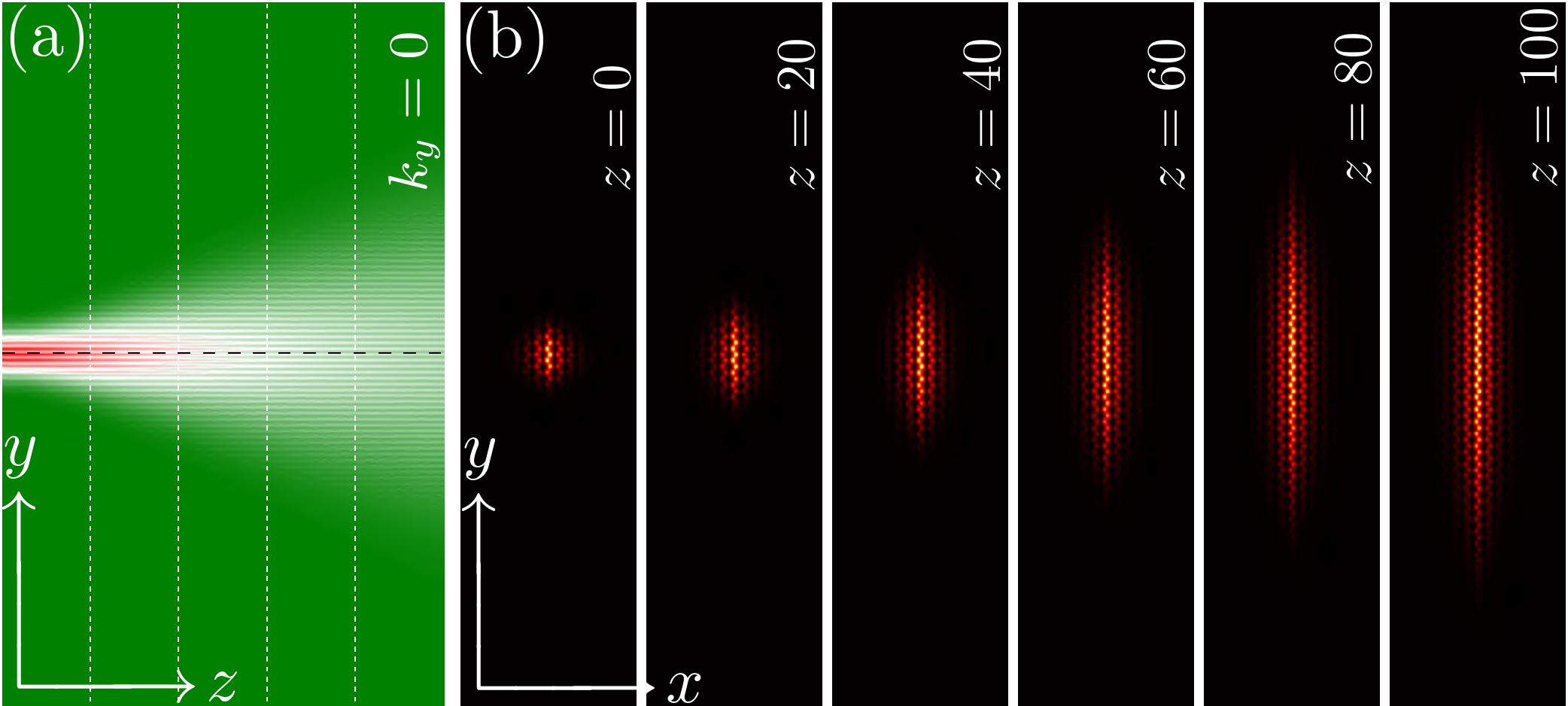}
\caption{(a) Propagation dynamics of the valley Hall edge state with 
	superimposed Gaussian envelope at $x=0$. The dashed line represents the trajectory of the beam center. (b) Field modulus distributions in the $(x,y)$ plane at different distances $z$ illustrating diffraction of such beam.}
\label{figs1}
\end{figure}

As one can see from the evolution dynamics in the $x=0$ cross-section, the beam shown in Fig.~\ref{figs1}(a) does not exhibit acceleration in the course of evolution. Its integral center
\[
\bar{y} = \frac{\displaystyle\iint y|\psi|^2 dxdy}{\displaystyle\iint |\psi|^2 dxdy},
\]
indicated by the horizontal dashed line in Fig.~\ref{figs1}(a), remains ${\bar{y}=0}$ during propagation, in clear contrast with evolution of self-accelerating beam in Fig.~\ref{fig2}(a). The field modulus distributions in the $(x,y)$ plane shown at different distances z in Fig.~\ref{figs1}(b) also reveal diffraction of the beam without the shift of its integral center.




\section{Self-accelerating valley Hall edge state in self-defocusing Kerr medium}   \label{appD} 

The nonlinear self-accelerating valley Hall edge states also exist in defocusing nonlinear Kerr medium. The envelope for such beams can be obtained by solving the ordinary differential equation
\begin{equation}\label{eqs2}
\frac{\partial ^2 w}{\partial \eta^2} - 2|w|^2w - 4\mu \left( \eta + \frac{b_{\rm nl}}{2\mu} \right) w=0,
\end{equation}
which can be obtained from the governing equation with defocusing Kerr nonlinearity
\begin{equation}
i\frac{\partial \psi}{\partial z} = - \frac{1}{2} \Delta \psi - \mathcal{R}(x,y) \psi + |\psi|^2 \psi,
\end{equation}
using the same procedure, as described in the main text.

\begin{figure}[h]
\centering
\includegraphics[width=\columnwidth]{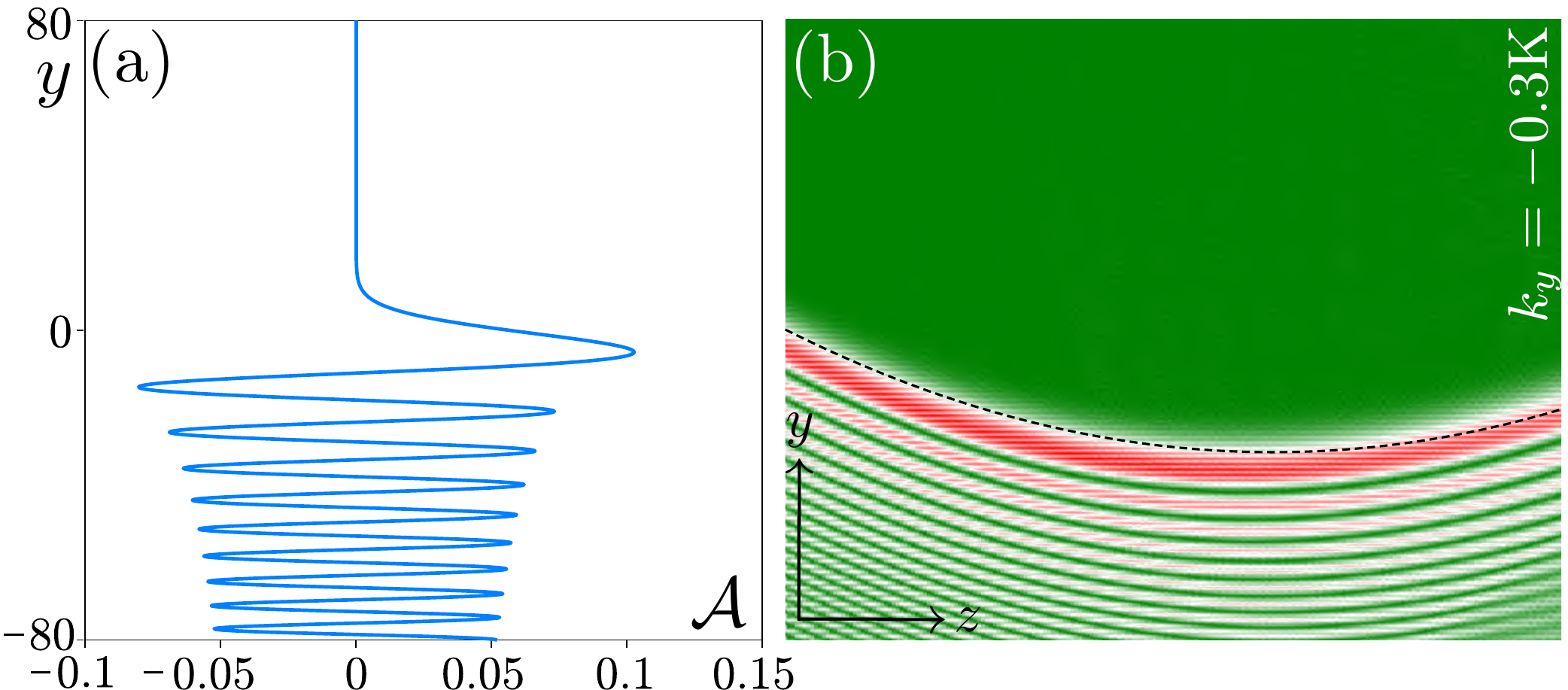}
\caption{(a) A nonlinear self-accelerating envelope. (b) Cross-section of the nonlinear self-accelerating valley Hall edge state during propagation in a self-defocusing nonlinear Kerr medium. The dashed curve is the predicted parabolic trajectory that is same as that in Fig.~\ref{fig4}(d). The panel is shown in ${-80\le y \le 80}$ and ${0\le z \le 200}$.}
\label{figs2}
\end{figure}

In Fig.~\ref{figs2}(a), we display an example of the envelope for such self-accelerating beam corresponding to ${|\mathcal{A}|_{\max}\sim0.1}$. By superimposing this envelope on the linear valley Hall edge state, the nonlinear self-accelerating valley Hall edge state is constructed. In Fig.~\ref{figs2}(b), we show the cross-section of the nonlinear self-accelerating valley Hall edge state with ${k_y=-0.3{\rm K}_y}$ during its propagation in defocusing medium. Just as in the case illustrated in Fig.~\ref{fig4}(d) in the main text, one observes that the beam changes its propagation direction upon evolution, while maintaining its internal structure.

\section*{Acknowledgment}
This work was supported by the Natural Science Basic Research Program of Shaanxi Province (2024JC-JCQN-06, 2025JC-QYCX-006) and the National Natural Science Foundation of China (12474337). Y.V.K. acknowledges funding by the Russian Science Foundation (grant 24-12-00167) and partially by the research project FFUU-2024-0003 of the Institute of Spectroscopy of the Russian Academy of Sciences.


%

\end{document}